\documentclass[final,8pt,5p,twocolumn]{elsarticle}

\usepackage{amssymb}

\usepackage[final]{changes}

\usepackage{forest}

\usepackage{ulem}

\usepackage{xcolor}

\usepackage{tabularx,ragged2e}
\newcommand{\HY}{\hyphenpenalty=25\exhyphenpenalty=25} 
\newcolumntype{Z}{>{\HY\RaggedRight\arraybackslash\hspace{0pt}}X} 

\usepackage[hyphens]{url}
\usepackage{hyperref} 
\hypersetup{breaklinks=true}
\usepackage{caption}
\usepackage{booktabs}
\usepackage{multirow}
\usepackage{tabularx}	
\usepackage{color,colortbl}
\usepackage{subcaption}
\usepackage{graphicx}  

\usepackage[smartEllipses]{markdown}

\usepackage{listings}
\lstdefinelanguage{JavaScript}{
  keywords={break, case, catch, continue, debugger, default, delete, do, else, finally, for, function, if, in, instanceof, new, return, switch, this, throw, try, typeof, var, void, while, with},
  morecomment=[l]{//},
  morecomment=[s]{/*}{*/},
  morestring=[b]',
  morestring=[b]",
  sensitive=true
}
\usepackage{xcolor}
\definecolor{codegreen}{rgb}{0,0.6,0}
\definecolor{codegray}{rgb}{0.5,0.5,0.5}
\definecolor{codepurple}{rgb}{0.58,0,0.82}
\definecolor{backcolour}{rgb}{0.95,0.95,0.92}

\lstdefinestyle{mystyle}{
    backgroundcolor=\color{backcolour},   
    commentstyle=\color{codegreen},
    keywordstyle=\color{magenta},
    numberstyle=\tiny\color{codegray},
    stringstyle=\color{codepurple},
    basicstyle=\ttfamily\footnotesize,
    breakatwhitespace=false,         
    breaklines=true,                 
    captionpos=b,                    
    keepspaces=true,                 
    numbers=left,                    
    numbersep=5pt,                  
    showspaces=false,                
    showstringspaces=false,
    showtabs=false,                  
    tabsize=2
}

\colorlet{punct}{red!60!black}
\definecolor{background}{HTML}{EEEEEE}
\definecolor{delim}{RGB}{20,105,176}
\colorlet{numb}{magenta!60!black}

\lstdefinelanguage{json}{
    numbers=left,
    numberstyle=\scriptsize,
    stepnumber=1,
    numbersep=8pt,
    breaklines=true,
    backgroundcolor=\color{background},
    literate=
     *{0}{{{\color{numb}0}}}{1}
      {1}{{{\color{numb}1}}}{1}
      {2}{{{\color{numb}2}}}{1}
      {3}{{{\color{numb}3}}}{1}
      {4}{{{\color{numb}4}}}{1}
      {5}{{{\color{numb}5}}}{1}
      {6}{{{\color{numb}6}}}{1}
      {7}{{{\color{numb}7}}}{1}
      {8}{{{\color{numb}8}}}{1}
      {9}{{{\color{numb}9}}}{1}
      {:}{{{\color{punct}{:}}}}{1}
      {,}{{{\color{punct}{,}}}}{1}
      {\{}{{{\color{delim}{\{}}}}{1}
      {\}}{{{\color{delim}{\}}}}}{1}
      {[}{{{\color{delim}{[}}}}{1}
      {]}{{{\color{delim}{]}}}}{1},
}

\makeatletter
\newcommand{\ion}[2]{%
  #1$\;$%
  \if b\expandafter\@car\f@series\relax\@nil
    \begingroup 
      \sbox0{\rmfamily\mdseries\textsc{v}}%
      \resizebox{!}{\ht0}{\rmfamily\@Roman{#2}}%
    \endgroup
  \else
    \textsc{\rmfamily\@roman{#2}}%
  \fi
}
\makeatother

\makeatletter
\def\@author#1{\g@addto@macro\elsauthors{\normalsize%
    \def\baselinestretch{1}%
    \upshape\authorsep#1\unskip\textsuperscript{%
      \ifx\@fnmark\@empty\else\unskip\sep\@fnmark\let\sep=,\fi
      \ifx\@corref\@empty\else\unskip\sep\@corref\let\sep=,\fi
      }%
    \def\authorsep{\unskip,\space}%
    \global\let\@fnmark\@empty
    \global\let\@corref\@empty  
    \global\let\sep\@empty}%
    \@eadauthor={#1}
}
\makeatother

\journal{Journal of Systems and Software}

\begin{document}

\begin{frontmatter}

\title{BugMagnifier: TON Transaction Simulator for Revealing Smart Contract Vulnerabilities}

\affiliation[Sk]{organization={Skolkovo Institute of Science and Technology},
            city={Moscow},
            country={Russia}}


\affiliation[PT]{organization={Positive Technologies},
            city={Moscow}, country={Russia}}

\affiliation[MSU]{organization={Lomonosov Moscow State University},
            city={Moscow}, country={Russia}}

\affiliation[MIPT]{organization={Moscow Institute of Physics and Technology},
            city={Moscow},
            country={Russia}}

\affiliation[IIT]{organization={Indian Institute of Technology},
            city={Delhi},
            country={India}}

\author[Sk]{Yury Yanovich}
\ead{{Corresponding author*}y.yanovich@skoltech.ru}

\author[MIPT]{Victoria Kovalevskaya}
\ead{vvkovalevskaya@phystech.edu}

\author[MIPT]{Maksim Egorov}
\ead{i@med2048.ru}

\author[MIPT]{Elizaveta Smirnova}
\ead{smirnova.elizaveta.s@phystech.edu}

\author[MIPT]{Matvey Mishuris}
\ead{mishuris.ma@phystech.edu}

\author[Sk]{Yash Madhwal}
\ead{yash.madhwal@skoltech.ru}

\author[PT,MSU]{Kirill Ziborov}
\ead{kziborov@ptsecurity.com}

\author[MIPT]{Vladimir Gorgadze}
\ead{gorgadze@gmail.com}

\author[IIT]{Subodh Sharma}
\ead{svs@iitd.ac.in}  

\begin{abstract}
The Open Network (TON) blockchain employs an asynchronous execution model that introduces unique security challenges for smart contracts. A primary concern is race conditions arising from unpredictable message processing order. While previous work established vulnerability patterns through static analysis of audit reports, dynamic detection of temporal dependencies through systematic testing remains an open problem. This study proposes a dynamic evaluation methodology based on controlled message orchestration to systematically expose vulnerabilities in asynchronous smart contracts. By synthesizing precise message queue manipulation with differential state analysis and probabilistic permutation testing, we establish a framework (namely, BugMagnifier) for identifying execution flaws that static methods miss. Experimental evaluation demonstrates BugMagnifier's effectiveness through extensive parametric studies on purpose-built vulnerable contracts and five real-world vulnerability cases reproduced from recent security audits. Results reveal message ratio-dependent detection complexity that aligns with theoretical predictions. This quantitative model enables predictive vulnerability assessment while shifting discovery from manual expert analysis to automated evidence generation. By providing reproducible test scenarios for temporal vulnerabilities, BugMagnifier addresses a critical gap in the TON security tooling, offering practical support for safer smart contract development in asynchronous blockchain environments.
\end{abstract}

\begin{keyword}
Blockchain security \sep smart contract \sep asynchronous systems \sep vulnerability detection \sep transaction simulation
\end{keyword}

\end{frontmatter}

\section{Introduction}
\label{sec:intro}
Blockchain architectures face an inherent tension between transactional atomicity and network scalability, a fundamental trade-off defining platform security profiles~\cite{Buterin2021a,Kruglik2019a,Amelin2021}. Ethereum's synchronous model~\cite{Buterin2014} demonstrates this through complex atomic operations such as decentralized exchange (DEX) arbitrage~\cite{Gramlich2024,Vostrikov2025}, enabling coordinated asset swaps within single transactions. These operations rely on deterministic execution ordering, ensuring static market conditions during processing. However, this atomicity restricts throughput to approximately 30 transactions per second (TPS), limiting scalability to single-chain performance.

The Open Network (TON) addresses scalability through a sharded architecture where transactions decompose into independent messages processed concurrently across chains~\cite{Durov2019,Berger2019}. This asynchronous model enables theoretical throughput exceeding 100,000 TPS, i.e. an improvement of more than three orders of magnitude. However, this innovation introduces novel security challenges. Where Ethereum implements arbitrage atomically, TON requires decomposition into interdependent messages. Developers must manage state changes between executions, cross-shard fees, and partial failure recovery. Unpredictable message interleaving creates distinct vulnerability patterns, with temporal dependencies becoming predominant.

TON's architecture inherently amplifies race condition risks through five-phase transaction processing, where intermediate states become visible to subsequent messages. Cell-based storage of TVM exacerbates risks through dynamic fee calculations based on the depth of the cell tree, enabling financial attacks via strategic message sequences~\cite{TON2025a}. Official TON security guidelines explicitly warn about race conditions in message flows and account destruction scenarios, categorizing them as critical vulnerabilities requiring preventive measures.

Whilst Ethereum's ecosystem has experienced critical vulnerabilities like The DAO's reentrancy attack and Parity's access control flaws~\cite{Atzei2017,Chen2022a,Soud2024}, these differ fundamentally from TON's architecturally induced race conditions. Ethereum's documented race condition vulnerabilities~\cite{Breidenbach2018} typically require deliberate transaction ordering rather than emerging from inherent design constraints.
The nascent TON security ecosystem currently offers analyzers like TONScanner~\cite{Song2025} and TSA~\cite{Espirito2025}, which are effective for identifying static code defects (e.g., bad randomness, unchecked returns). However, these tools operate via static analysis and symbolic execution, methodologies that are inherently unable to simulate the temporal execution paths arising from non-deterministic message interleaving. Consequently, a critical gap exists for detecting dynamic state-dependent race conditions, leaving developers without systematic methods to validate contract behavior under asynchronous execution.

This paper addresses this critical need by introducing BugMagnifier, a transaction scheduling and simulation framework designed specifically to reveal vulnerabilities in TON smart contracts through controlled message orchestration. Unlike conventional analysis tools, BugMagnifier operates at the message granularity level, enabling precise manipulation of message queues and TVM states to systematically expose temporal dependencies. The framework implements a novel permutation testing methodology that quantifies vulnerability manifestation probability based on message ratios, addressing the core challenge of non-deterministic message ordering in TON's architecture. The key contributions of this paper are as follows:
\begin{enumerate}
\item \textbf{Designed and developed the TON-Complete Transaction Simulation} that integrates with TON Sandbox to enable local emulation of the TON's complete transaction processing cycle, including all five execution phases, 
a capability absent in existing tools that typically support only the compute phase.
\item \textbf{Implemented a message permutation engine} that systematically explores execution paths through configurable message ordering, enabling detection of race conditions that would remain undetected by conventional testing methodologies. The engine incorporates probabilistic analysis to identify vulnerable message ratios and quantify detection complexity.
\item \textbf{Established a quantitative model} of race condition manifestation in TON smart contracts, validated through extensive execution permutations on a purpose-built vulnerable contract and five real-world vulnerability patterns reproduced from recent security audits. Our experiments reveal a message ratio-dependent detection complexity that aligns with theoretical predictions, providing actionable insights for security testing protocols.
\end{enumerate}

By shifting vulnerability discovery from manual expert analysis to systematic evidence generation, BugMagnifier addresses a critical gap in TON security tooling. The framework provides reproducible test scenarios for asynchronous execution flaws, particularly race conditions, that have proven exceptionally challenging to detect through conventional methods. Our work contributes to a more secure and robust blockchain environment by empowering developers with practical tools to build safer and more reliable decentralized applications on TON, ultimately supporting the ecosystem's maturation and adoption.

The remainder of this paper is organized as follows. Section~\ref{sec:Background} provides a technical foundation of TON's architecture with a specific focus on the security implications of its asynchronous execution model. Section~\ref{sec:RelatedWork} analyzes the limitations of existing security tools in addressing TON's unique vulnerability landscape. Section~\ref{sec:ProposedSolution} details BugMagnifier's design, implementation, and technical innovations. Section~\ref{sec:Example} demonstrates the capabilities of the framework through a deliberately vulnerable smart contract. Section~\ref{sec:NumericalExperiments} presents an empirical validation of our approach through comprehensive numerical experiments. Finally, Section~\ref{sec:Conclusion} discusses implications for TON security practices and future research directions.

\section{Background}
\label{sec:Background}

\subsection{TON Virtual Machine (TVM)} 

The TON Virtual Machine (TVM) serves as the deterministic execution environment for smart contracts on The Open Network, operating as a stack-based virtual machine with specific architectural features tailored to TON's sharded blockchain design~\cite{Durov2020}. As blockchains operate in an untrusted environment where participants must self-verify computational consistency, computation results need to be hardware-independent; virtual machines like TVM are the standard choice to ensure this determinism across all network nodes. Unlike Ethereum's virtual machine (EVM), which processes transactions as atomic units, TVM is specifically engineered to handle TON's message-centric execution model, where transactions decompose into independent messages processed across (parallel) shard chains. The TVM state comprises five critical components: Stack (for data manipulation), Control Registers (managing execution flow), Current Continuation (tracking program state), Current Codepage (storing executable code), and Gas Limits (enforcing computational constraints).

To prevent deadlock and ensure liveness, the user program execution in TVM, as in EVM, is strictly limited in both computation and memory. This limitation is expressed in universal gas units. The protocol pre-defines a cost in gas for every elementary operation and for storage, forming the basis for transaction fees that users are charged based on the gas consumed by their transactions.

TVM implements a comprehensive instruction set optimized for blockchain operations, featuring 256 primitive opcodes organized into logical groups. The stack manipulation primitives include basic operations like \texttt{PUSH}, \texttt{POP}, and \texttt{XCHG}, as well as compound operations such as \texttt{PUXC} and \texttt{XCHG3} that enhance code density. For data handling, TVM utilizes cells as its fundamental data structure, recursive tree-like containers that store serialized data in bit strings and references to other cells. This cell-based architecture enables efficient Merkle tree constructions essential for TON's sharding mechanism. TVM's cell operations include specialized primitives like \texttt{STORING} (for storage fee calculation) and \texttt{ACCEPT\_MSG\_VALUE} (for credit phase processing), which directly impact security considerations in smart contract development.

The TVM instruction set incorporates blockchain-specific functionality through application-oriented primitives, including message handling (\texttt{SEND\_RAW\_MESSAGE}), gas management (\texttt{ACCEPT\_MSG\_VALUE}), and cryptographic operations (\texttt{SHA256\_C}). Notably, TVM implements a five-phase transaction processing model: storage, credit, compute, action, and bounce, each with distinct gas accounting rules and state transition semantics. This multi-phase execution model creates unique security challenges, as vulnerabilities can manifest across phase boundaries rather than within isolated transaction contexts. The compute phase, where contract logic is executed, operates under strict gas limits that vary based on preceding phase outcomes, requiring precise gas calculations to avoid out-of-gas failures that could leave contracts in inconsistent states.

This phased gas model within TVM is a direct consequence of, and is designed to support, TON's larger asynchronous message-passing architecture.

\subsection{Asynchronous Transaction Execution}  

TON's asynchronous execution model fundamentally differs from conventional blockchains through its message-driven architecture, where transactions decompose into independent messages processed concurrently across shard chains. Each message undergoes five sequential processing phases: storage (calculating storage fees), credit (accounting for incoming funds), compute (executing contract logic), action (generating new messages) and bounce (handling failed message processing). This phased execution enables higher throughput by allowing parallel processing of independent messages while maintaining deterministic outcomes through the consensus algorithm that establishes a final total order for messages.

The asynchronous model introduces critical security considerations that are absent in synchronous blockchains. While the consensus algorithm ensures a final deterministic outcome across the network, the concurrent processing of messages across shards creates a period of temporal non-determinism from the perspective of a single smart contract. The order in which messages from different senders interleave in their queue is not fixed until included in a block, creating race conditions when multiple messages target the same contract state. Unlike Ethereum's transaction atomicity, where state changes occur only upon successful transaction completion, TON contracts experience intermediate state transitions after each message processing phase, particularly during the compute phase, which can be observed by subsequent messages. This behavior enables temporal vulnerabilities where the contract state appears consistent during individual message processing but becomes inconsistent when messages interleave unexpectedly.

The bounce phase represents a unique TON security consideration, where failed message processing triggers automatic fund recovery through bounce messages. Contracts must implement proper bounce message handling using the \texttt{SET\_BOUNCE} primitive; failure to do so can result in permanent fund loss during message processing failures. Additionally, TON's storage fee mechanism, calculated dynamically during the storage phase based on cell tree depth, creates financial attack vectors where malicious actors could force contracts into storage debt through carefully crafted message sequences.

A significant emerging trend is the application of Artificial Intelligence and Large Language Models (LLMs) for automated vulnerability discovery and exploit generation. Recent studies demonstrate that models like GPT-4 and Claude can analyze Solidity code and generate functional proof-of-concept exploits for known vulnerability patterns with notable success rates~\cite{Xiao2025}. Although these approaches excel at reasoning about code semantics and predicting flaws based on learned patterns, they operate through static analysis or limited symbolic execution. BugMagnifier represents a complementary paradigm: instead of predicting vulnerabilities from code, it concretely executes contract logic under systematically varied message orderings to discover temporal flaws. This dynamic exploration-based approach is crucial for detecting asynchronous race conditions that are often poorly represented in the training data of current LLMs.

\subsection{Comparative Analysis of Execution Models}  

The architectural divergence between TON and Ethereum creates a significant conceptual challenge for developers transitioning between ecosystems. In Ethereum's synchronous model, transactions execute atomically within blocks, ensuring that intermediate states remain invisible to external contracts until finalization. This guarantees temporal consistency, where state changes occur only upon successful transaction completion. Conversely, TON's asynchronous model exposes intermediate contract states after each message processing phase, requiring developers to design contracts that maintain consistency across partial execution states.

The most critical translation challenge involves concurrency handling: Ethereum developers typically assume sequential transaction execution, while TON developers must explicitly account for message interleaving. For example, an Ethereum contract implementing a simple token transfer would execute as a single atomic operation, whereas the equivalent TON implementation requires careful state management across multiple messages, deposit, validation, and withdrawal, each potentially interleaved with messages from other users. This fragmentation necessitates explicit synchronization mechanisms, such as nonce-based state validation or message sequencing constraints, which have no direct Ethereum equivalents.

Another fundamental difference lies in error handling semantics. Ethereum transactions revert entirely upon failure, while TON's phased execution allows partial state changes to persist even when subsequent phases fail. The bounce phase specifically handles message processing failures by generating recovery messages, requiring contracts to implement defensive patterns that prevent state inconsistencies when bounce messages interleave with new operations. These architectural differences necessitate a complete rethinking of security patterns, as conventional Ethereum security practices like reentrancy guards become insufficient for addressing TON's message interleaving vulnerabilities.

\subsection{Implications for Smart Contract Development}  

The asynchronous execution model of TON necessitates fundamentally different development practices compared to synchronous blockchains. Developers must implement explicit state validation at each message processing stage, as contracts can be invoked with partially updated states from previous incomplete transactions. This requirement transforms seemingly simple operations, like balance transfers, into complex state machines where intermediate states must remain consistent under arbitrary message interleaving.

Security-critical contracts require sophisticated message sequencing constraints to prevent temporal vulnerabilities. For example, deposit-withdrawal patterns must incorporate nonce mechanisms or timestamp validation to ensure message processing order integrity, as the natural message queue ordering provides no guarantees of sequential execution. The storage fee mechanism further complicates development, as contracts must maintain sufficient balance not only for operational gas but also for dynamic storage costs that vary with cell tree depth, a vulnerability vector where attackers can force contracts into storage debt through strategically crafted messages.

TON's bounce message system introduces additional security considerations, as contracts must properly handle both successful and bounced message processing paths. Failure to implement comprehensive bounce message validation can lead to fund loss or state corruption when processing interleaved message sequences. The phased transaction model also requires precise gas budgeting across all five phases, as gas exhaustion at any stage produces different failure modes, from partial state updates to complete transaction rollback, depending on the failure phase.

These complexities necessitate specialized security testing approaches that explicitly model message ordering permutations, as conventional test methodologies designed for atomic transactions cannot adequately validate contract behavior under TON's asynchronous execution model. The lack of temporal determinism in message processing requires developers to adopt formal verification techniques or systematic permutation testing to ensure contract resilience against race conditions and other temporal vulnerabilities.

\section{Related Work}
\label{sec:RelatedWork}

The security of smart contracts has matured into a significant research domain, though its focus remains predominantly on the Ethereum Virtual Machine (EVM) and its synchronous execution model. This section reviews the foundational work in smart contract vulnerability analysis, the tooling built for the EVM-centric paradigm, and the emerging research on transaction ordering in adversarial environments. We then contextualize existing TON-specific analysis within this broader landscape, concluding with a comparative analysis that highlights the gap filled by our work: a dynamic transaction simulator for the asynchronous TON blockchain.

\subsection{General Smart Contract Vulnerabilities and Empirical Studies}
\label{subsec:vuln-studies}

The systematic study of smart contract vulnerabilities was pioneered by researchers analyzing EVM and Solidity. Early work established taxonomies of common flaws, such as reentrancy and integer overflows, which have become the primary targets for automated audit tools. This foundation is crucial, yet the specific vulnerability patterns are often tied to the semantics of the EVM opcodes and Solidity's execution model.

Recent empirical studies continue to deepen this understanding by analyzing large-scale contract deployments. For example, Chen et al.~\cite{Chen2019b} conducted a large-scale empirical study focusing on the critical task of identifying control flow in smart contracts, a foundational step for many advanced static analyzes. Their work on Ethereum contracts highlights the methodologies and challenges of reasoning about contract execution paths, which differs significantly in TON's message-passing architecture. Furthermore, Li et al.~\cite{Li2025a} provided a contemporary empirical analysis of \emph{inconsistent state update} vulnerabilities. While their study is grounded in the EVM paradigm, the core concept--unexpected state interleaving due to concurrent operations--is directly analogous to the race conditions endemic to TON's asynchronous model, underscoring the universal nature of state consistency challenges in decentralized systems.

\subsection{Analysis Tools for the EVM and Synchronous Paradigms}
\label{subsec:evm-tools}

The maturity of Ethereum security research is evidenced by a robust ecosystem of analysis tools employing static, dynamic, and formal methods. Static analysis frameworks like Slither~\cite{Feist2019} have set a high standard; by converting Solidity source code into an intermediate representation and applying data-flow and taint analysis, they efficiently detect dozens of common vulnerability patterns without execution. Automated invariant generation techniques \cite{Liu2025a} further enhance this by inferring and checking the properties that must hold across all executions, providing a higher level of assurance for synchronous contract models. Similarity-based analysis \cite{Chen2025} offers another angle by comparing contract codebases to known vulnerable patterns, efficiently scaling to large ecosystems. Dynamic analysis and fuzzing tools, such as Echidna \cite{Grieco2020} and xFuzz~\cite{Xue2024}, take a complementary approach by executing contracts with generated or mutated inputs to uncover unexpected runtime behavior. Dynamic exploit generation \cite{Wang2022c} advances this further by automatically creating concrete proofs-of-concept for detected vulnerabilities. On the formal verification front, tools like the Certora Prover represent the high-assurance end of the spectrum, using mathematical proofs to guarantee that a contract's implementation satisfies its formal specification. Similarly, correct-by-design approaches such as VeriSolid \cite{Nelaturu2023} enable the specification and verification of interacting contracts' state machines, offering strong guaranties for synchronous systems like Ethereum.

Whilst powerful, these tools are architected around the assumptions of the EVM: atomic transaction execution, a globally ordered block sequence, and synchronous state transitions. They reason about a single transaction's effect on a contract's state in isolation. Consequently, they lack the fundamental primitives to model TON's core challenges: the non-deterministic interleaving of \emph{multiple independent messages} targeting the same contract and the state visibility across the five distinct processing phases (storage, credit, compute, action, bounce). A tool like Slither cannot analyze the flow of messages between accounts, and a fuzzer like Echidna cannot systematically permute the order of a message queue to discover race conditions.

\begin{table*}[t]
\centering
\caption{Comparative Analysis of Smart Contract Security Approaches}
\label{tab:comparative-analysis}
\begin{tabularx}{\textwidth}{@{}>{\raggedright\arraybackslash}p{2.2cm} >{\centering\arraybackslash}p{1.6cm} >{\centering\arraybackslash}p{1.8cm} >{\centering\arraybackslash}p{1.4cm} >{\raggedright\arraybackslash}p{4.0cm} >{\raggedright\arraybackslash}p{4.8cm}@{}}
\toprule
\textbf{Tool / Approach} & \textbf{Target Platform} & \textbf{Core Technique} & \textbf{Models Async. Queues?} & \textbf{Primary Strength} & \textbf{Limitation for TON Asynchronous Vulnerabilities} \\
\midrule
\textbf{Slither} \cite{Feist2019} & EVM & Static Analysis & \textbf{\texttimes} & Fast detection of code pattern vulnerabilities in Solidity. & EVM-centric; cannot model TON's message lifecycle or cell semantics. \\
\addlinespace
\textbf{Echidna} \cite{Grieco2020} & EVM & Fuzzing & \textbf{\texttimes} & Property-based testing via execution with random inputs. & Generates linear transaction sequences, not interleaved, concurrent messages. \\
\addlinespace
\textbf{MEV Research} \cite{Daian2020} & EVM & Economic Analysis & \textbf{*} & Models profit from adversarial transaction ordering. & Focuses on miner economics, not contract-state race condition detection. \\
\addlinespace
\textbf{VeriSolid} \cite{Nelaturu2023} & EVM & Formal Verification / Model Checking & \textbf{\texttimes} & Correct-by-design construction and verification of interacting state machines. & Designed for synchronous atomic transactions; cannot model TON's message lifecycle or interleavings. \\
\addlinespace
\textbf{Concurrency Testing} \cite{Elmas2013,Terragni2025} & Generic Software & Dynamic Analysis & \textbf{\checkmark} (concepts) & Systematic exploration of thread/interleaving bugs. & Framework not adapted for blockchain message-passing or TVM state. \\
\addlinespace
\textbf{TONScanner} \cite{Song2025} & TON & Static Analysis & \textbf{\texttimes} & Detects 8+ TON-specific code defect patterns in FunC. & No execution; cannot simulate message queues or detect ordering flaws. \\
\addlinespace
\textbf{TSA} \cite{Espirito2025} & TON & Symbolic Execution & \textbf{\texttimes} & Finds runtime errors & Path explosion limits exploration of multi-message permutations. \\
\midrule
\textbf{BugMagnifier (Proposed)} & \textbf{TON} & \textbf{Dynamic Simulation \& Permutation Testing} & \textbf{\checkmark} & \textbf{Systematic race condition detection via controlled message orchestration.} & \textbf{N/A (Designed to address the above limitations)} \\
\bottomrule
\end{tabularx}
\vspace{2mm}
\footnotesize{* Analyzes ordering within a synchronous block, not an asynchronous message queue.}
\end{table*}

\subsection{Adversarial Transaction Ordering and MEV Research}
\label{subsec:mev-research}

Research into Maximum Extractable Value (MEV) has extensively explored the adversarial manipulation of transaction ordering within a block, a concern that shares a conceptual lineage with TON's race conditions. The seminal work by Daian et al.~\cite{Daian2020} in ``Flash Boys 2.0'' documented how bots can front-run, back-run, or sandwich transactions in Ethereum's mempool to extract profit, leading to consensus instability and economic centralization. This body of work formally treats the blockchain state as a function of an ordered transaction list and examines the consequences of perturbing that order.

Our work is philosophically aligned with this research in its focus on \emph{ordering-dependent state outcomes}. However, a critical distinction exists: MEV research primarily studies the \emph{economic} exploitation of ordering in a public mempool for profit within a \emph{synchronous} block. In contrast, TON's vulnerability is \emph{architectural}: the lack of atomicity means that even two honest, correctly formatted messages can create a financially damaging race condition based solely on their arrival order, regardless of miner/validator intent. BugMagnifier applies the core investigative principle--systematically exploring ordering permutations--to this different, platform-inherent security problem.

\subsection{Foundations in Concurrent System Verification}
\label{subsec:concurrent-systems}

The core challenge of detecting state-dependent race conditions in TON is conceptually analogous to the classic problem of verifying correctness in concurrent software, where non-deterministic thread scheduling can lead to Heisenbugs. Research in this domain has developed robust methodologies for systematic exploration. Frameworks like CONCURRIT introduced programmable thread interleaving for testing, allowing developers to specify and exhaustively explore specific scheduling scenarios of interest~\cite{Elmas2013}. Similarly, research on differential testing of concurrent classes provides formal methods for identifying behavioral divergences between different implementations under all possible interleavings~\cite{Terragni2025}.

BugMagnifier adapts this foundational principle to the blockchain domain. It replaces the abstraction of thread scheduling with the concrete, controlled manipulation of TON's message queue. While traditional tools reason about shared memory and locks, BugMagnifier operates on TVM state transitions and cell semantics. This connection positions our work not as an ad hoc TON-specific solution, but as a principled application of established concurrent system verification techniques to the novel execution model of asynchronous blockchains.

\subsection{Prior Work on TON Smart Contract Security}
\label{subsec:ton-prior-work}

Security analysis for TON is a nascent field. Initial efforts have rightly focused on porting established techniques to the new platform. Song et al.'s TONScanner~\cite{Song2025} represents a critical first step, applying static analysis patterns to FunC source code to identify coding errors like unchecked bounced messages or precision loss. Similarly, the Ton Symbolic Analyzer (TSA)~\cite{Espirito2025} employs symbolic execution on TVM bytecode to find runtime errors. These tools are invaluable for detecting \emph{intra-transaction} and \emph{single-message} bugs, analogous to the first generation of Ethereum tools.

For Ethereum, formal verification tools like Certora Prover represent the high-assurance end of the security spectrum, using mathematical proofs to guarantee that a contract's implementation satisfies its formal specification~\cite{Certora2025}. While such tools provide the strongest guaranties for synchronous blockchains, their adaptation to TON's asynchronous model is an open challenge. Verifying temporal properties--such as the absence of race conditions across all possible message interleavings--requires sophisticated modeling of the message queue and TVM phases, a problem analogous to the formal verification of distributed, concurrent systems. BugMagnifier's permutation testing can be viewed as a dynamic and practical stepping stone towards such formal guaranties, helping developers identify flawed contracts and gather concrete execution traces that could later inform the creation of formal models and specifications.

Their limitation, as highlighted in our introduction, is their inability to address the \emph{inter-message} and \emph{temporal} vulnerabilities that define TON's unique threat model. Static analysis cannot simulate the execution of a message queue, and symbolic execution becomes computationally intractable when modeling all possible permutations of even a small set of concurrent messages. This creates a gap where a contract can pass all existing TON linters and verifiers, yet remain highly vulnerable to race conditions.

\subsection{Comparative Analysis and Identified Gap}
\label{subsec:comparative-gap}

As illustrated in Table~\ref{tab:comparative-analysis}, existing tools are bifurcated into two categories: 1) mature solutions for a synchronous paradigm that cannot model TON architecture, and 2)~nascent TON-specific tools that cannot analyze the temporal dependencies arising from concurrency. Research on transaction ordering, while insightful, does not provide a practical testing framework for TON smart contracts.

\textbf{BugMagnifier} is designed to bridge this gap directly. It takes the adversarial ordering exploration principle from MEV research and implements it as a practical, dynamic analysis tool for the TON blockchain. By providing granular control over the TVM's execution phases and the message queue, it enables the systematic discovery of vulnerabilities that are invisible to all other approaches in the TON security toolkit. Our work thus advances the field by adapting a core security analysis methodology--permutation testing--to the unique challenges of asynchronous, sharded blockchain platforms.

\section{Proposed Solution}
\label{sec:ProposedSolution}

This section provides a comprehensive description of BugMagnifier, a specialized framework designed to detect and demonstrate vulnerabilities in smart contracts within the asynchronous transaction model of the TON blockchain. Unlike conventional blockchain analysis tools that operate at the transaction level, BugMagnifier introduces message-level execution control, a critical capability for exposing temporal vulnerabilities that manifest only under specific message interleaving sequences. The framework's architecture centers on precise manipulation of message processing order while maintaining complete visibility into contract state transitions, enabling systematic exploration of execution paths that would remain hidden in conventional testing approaches. This methodology addresses the fundamental challenge of non-deterministic message ordering in the TON's architecture, where vulnerability manifestation depends on the sequence of message processing rather than isolated transaction properties.

\subsection{Choosing a Blockchain Emulator}

The foundation of BugMagnifier's effectiveness lies in its ability to model TON's complete transaction processing mechanics with surgical precision. To achieve this, the emulator must satisfy three non-negotiable requirements derived from TON's architectural constraints: 
\begin{enumerate}
    \item comprehensive support for all five transaction phases (storage, credit, compute, action, and bounce),
    \item access to intermediate TVM states between phase transitions,
    \item extensibility to manipulate message queue ordering.
\end{enumerate}

These requirements emerged from the recognition that temporal vulnerabilities in TON contracts often manifest across phase boundaries rather than within isolated execution contexts.

Our evaluation revealed critical limitations in the existing tools. TON Contract Executor~\cite{TONContractExecutor} abstracts away all phases except compute, bypassing essential TVM opcodes such as \texttt{ACCEPT\_MSG\_VALUE} and \texttt{STORING}. This creates dangerous blind spots for vulnerabilities involving storage fees or bounce message handling. Similarly, the TVM-linker~\cite{TVMLinker} implements storage fees as fixed overhead rather than through TON's dynamic cell depth calculations and omits bounce phase processing entirely. Both tools fail to preserve intermediate TVM states between phase transitions, rendering them incapable of detecting state inconsistencies that occur during partial execution sequences.

TON Sandbox~\cite{Sandbox} emerged as the only viable foundation due to its transparent implementation of TON's transaction lifecycle. Its transaction processor module exposes the complete TVM state machine with precise gas accounting at each opcode execution, preserving the full context of intermediate states between phase transitions. Crucially, its modular architecture provides hooks into the message processing pipeline, enabling BugMagnifier's core innovation: a message interceptor that dynamically reorders messages while maintaining visibility into all five transaction phases. This capability allows systematic exploration of message sequences that trigger temporal vulnerabilities, particularly race conditions, where contract behavior depends on the interleaving of messages from multiple senders.

Building on TON Sandbox, BugMagnifier implements three critical enhancements: (1) a state snapshotting mechanism capturing complete TVM register states at each execution phase with minimal overhead (12\%); (2) a differential analyzer comparing memory layouts at the cell level to detect subtle inconsistencies; and (3) a message timing simulator modeling network latency effects on message interleaving. These innovations enable precise reproduction of edge cases where vulnerabilities manifest only under specific network conditions, such as storage debt scenarios triggered by strategically crafted cell tree depths or race conditions dependent on message arrival timing.

When initialized, BugMagnifier creates a local blockchain instance with capabilities exceeding standard development environments. The framework implements a copy-on-write state preservation system allowing rollback to any execution point, granular gas consumption control to simulate partial phase completions, and comprehensive execution tracing that maps message sequences to specific state transitions across all transaction phases. These features establish BugMagnifier as the first practical solution for systematic detection of temporal vulnerabilities in TON smart contracts, a class of flaws 
yet undetectable by conventional analysis methods.

\subsection{BugMagnifier}

The BugMagnifier interactive console is a powerful tool that provides developers with the ability to emulate TON transactions locally and identify vulnerabilities in smart contracts through advanced message management and simulation of various execution scenarios. The tool is publicly available on GitHub~\cite{BugMagnifier}. Built as a TypeScript application leveraging Node.js runtime, BugMagnifier implements a layered architecture with distinct components for contract compilation, transaction simulation, state management, and analysis. At its core, the tool establishes a bidirectional communication channel with the TON Sandbox through a custom IPC (Inter-Process Communication) protocol that enables fine-grained control over transaction execution while maintaining compatibility with the TVM's deterministic execution requirements.

During initialization, users specify the path to the target smart contract. Behind the scenes, BugMagnifier compiles the source code into executable TVM bytecode, saving the compiled code as a JSON file with base64 cell representation in the temporary directory \texttt{tmp/tondebug.compiled.json}. The compilation process utilizes the official FunC compiler with customized error reporting that maps bytecode offsets to source code locations, enabling precise debugging of complex contracts. Users can optionally configure the initial smart contract state and message queue before console launch, though both remain editable during interactive sessions. The initialization sequence implements a verification step that checks the contract's compliance with TON's compute phase limits, pre-emptively identifying potential out-of-gas conditions before execution begins.

The smart contract state configuration uses a JSON structure with three fields: contract \texttt{balance}, hex representation of TVM \texttt{code}, and base64-serialized contract state \texttt{data}. The balance parameter is particularly useful for simulating financial scenarios like insufficient balance handling during message processing or storage payments. Example state configuration:
\begin{lstlisting}[language=json,firstnumber=1,xleftmargin=3.0ex]
{
  "balance": "1099088800",
  "data": "b5ee9c...240000438014ffc...",
  "code": "b5ee9c...30000114Ff00f4a..."
}
\end{lstlisting}

The message queue employs a JSON array where each message contains these fields: unique identifier \texttt{id}, auto-generated if unspecified, internal/external-in \texttt{type}, defaults to internal, payload \texttt{body} interpreted by contract logic, nanotoken transfer amount \texttt{value} for internal messages, numerical sender identifier \texttt{senderId}, random if omitted, and optional test case label \texttt{name}.
Example message queue:
\begin{lstlisting}[language=json,firstnumber=1,xleftmargin=2.0ex]
[
  {
    "id": 1,
    "type": "internal",
    "body": "te6ccgEBAQEABgAACAAAAAI=",
    "value": {"coins": "10000000"},
    "senderId": 1,
    "name": "CLAIM Alice"
  },
  {
    "id": 2,
    "type": "internal",
    "body": "te6ccgEBAQEABgAACAAAAAE=",
    "value": {"coins": "10000000"},
    "senderId": 2,
    "name": "ENLIST Bob"
  }
]
\end{lstlisting}

Execution begins with the command:
\begin{lstlisting}[xleftmargin=2.0ex]
tondebug , contract <path> [, init-state <path>] 
[, queue <path>] [, help]
\end{lstlisting}

The console provides comprehensive message queue control through commands in the table \ref{sec:Table}.

\begin{table}[]
\centering
\footnotesize
\setlength{\tabcolsep}{4pt}
\caption{BugMagnifier Command Reference}
\label{tab:bugmagnifier_commands}
\begin{tabular}{|p{2.2cm}|p{6cm}|}
\hline
\textbf{Command} & \textbf{Functionality} \\
\hline
\texttt{run next} & Processes next message in queue, recording final state, 
transaction details, and message content \\
\hline
\texttt{run message <id>} & Executes message by ID (errors on invalid IDs) \\
\hline
\texttt{continue} & Processes all remaining messages \\
\hline
\texttt{queue list} & Displays current message queue \\
\hline
\texttt{set queue , order} & Modifies execution order (\texttt{reverse} or \texttt{random}) \\
\hline
\texttt{add messages <path>} & Adds new messages via JSON file \\
\hline
\texttt{delete message <id>} & Removes message by ID \\
\hline
\texttt{script load <path>} & Loads TS/JS script with \texttt{modifyQueue()} \\
\hline
\texttt{script run} & Applies loaded script to queue \\
\hline
\texttt{show state} & Displays contract status \\
\hline
\texttt{show transactions} & Shows transaction log \\
\hline
\texttt{show message log} & Displays message history \\
\hline
\texttt{load state <path>} & Loads new contract state \\
\hline
\texttt{save state <path>} & Saves current state to JSON \\
\hline
\texttt{diff <path1> <path2>} & Compares contract states \\
\hline
\texttt{exit} & Terminates console \\
\hline
\texttt{help} & Displays documentation \\
\hline
\end{tabular}\label{sec:Table}
\end{table}

Scripting examples demonstrate queue modification capabilities:

\textbf{JavaScript randomization script:}
{\small
\begin{lstlisting}[language=JavaScript,xleftmargin=2.0ex]
/**
@param {import('../tondebug/tondebug.ts')
.Message[]} queue
*/
export function modifyQueue(queue) {
for (let i = queue.length - 1;
i > 0; i, ) {
const j = Math.floor(
Math.random() * (i + 1));
[queue[i], queue[j]] =
[queue[j], queue[i]];
}
}
\end{lstlisting}
}
\emph{This script randomizes the message execution order using the Fisher-Yates shuffle algorithm to simulate non-deterministic race conditions and uncover ordering-dependent vulnerabilities.}

\textbf{TypeScript prioritization script:}
{\small
\begin{lstlisting}[language=JavaScript,xleftmargin=2.0ex]
export function modifyQueue(
queue: Message[]): void {
queue.sort((a, b) => {
const byType =
typePriority[a.type] -
typePriority[b.type];
if (byType !== 0) return byType;
const coinsA = a.value?.coins ?? 0n;
const coinsB = b.value?.coins ?? 0n;
return coinsB > coinsA ? 1 : coinsB < coinsA ? -1 : 0;
});
}
\end{lstlisting}
}
\emph{This script prioritizes messages by first comparing their type (e.g., external vs. internal) and then by descending value (\texttt{coins}), enabling testing of scenarios where message financial value or type determines processing priority.}

The scripting interface implements a secure sandbox environment that restricts access to system resources while providing full access to the message queue structure. This design enables complex test scenarios without compromising system security. Advanced users can implement sophisticated vulnerability detection algorithms through the scripting API, including probabilistic analysis of state transitions and custom anomaly detection heuristics.

The tool provides comprehensive control over smart contract testing by combining precise execution management (launching, removing, or adding specific messages) with automation capabilities (processing entire message queues) and thorough state analysis. This system enables: (1) processing order control through configurable message queues and scriptable race condition simulations; (2) change visualization via execution history and queue monitoring; and (3) systematic recording and analysis of all contract state changes post-transaction,  a critical capability for vulnerability detection that includes simulating complex attacks and assessing their impact on contract state. The differential analysis engine implements a cell-level comparison algorithm that identifies subtle state inconsistencies that are invisible to higher-level inspections, such as differences in cell bit patterns or hash tree structures. This capability is particularly valuable for detecting vulnerabilities involving complex data structures where surface-level state comparisons might miss critical discrepancies.

The solution supports the complete testing lifecycle, from environment initialization (automated FunC to TVM bytecode compilation and JSON-configured state setup) to post-execution analysis (including state comparison and race condition detection). Performance optimizations include a state caching mechanism that reduces redundant computation during repeated executions, and a parallel execution mode that leverages multi-core processors to accelerate permutation testing. Of particular value is the interactive mode, which allows dynamic message queue manipulation, execution order modification (sequential, reverse, or randomized), and integration of custom scripts for specialized test scenarios. The tool's architecture implements a plugin system that enables extension of core functionality without modifying the base code, facilitating integration with external analysis tools and continuous integration pipelines.

This combination of features establishes BugMagnifier as an essential tool for verifying smart contract's reliability and security prior to deployment on the live TON blockchain network. 
The ability to precisely control execution parameters and capture complete state transitions makes BugMagnifier particularly effective at identifying temporal vulnerabilities that would be nearly impossible to detect through conventional testing methods.

\section{Smart Contract Vulnerability Examples}
\label{sec:Example}

We now demonstrate BugMagnifier's vulnerability detection capabilities through a series of concrete examples. First, we present a synthetic smart contract system deliberately designed to exhibit a classic race condition; this controlled example serves as the foundation for the quantitative experiments in Section~\ref{sec:NumericalExperiments}. Subsequently, we introduce five real‑world vulnerabilities drawn from recent security audits, illustrating how BugMagnifier can uncover the same classes of temporal flaws in production‑grade contracts. Together, these examples validate both the pedagogical clarity and practical applicability of our approach.

\subsection{Synthetic Race Condition Example}

We employ BugMagnifier to demonstrate the detection of race conditions inherent to TON's asynchronous execution model. For this analysis, we designed smart contract system implementing a deposit pool with a critical vulnerability that manifests under concurrent message processing. This example illustrates how the non-deterministic ordering of messages in TON can lead to severe financial inconsistencies. The vulnerability originates from non-atomic state management during ownership assignment, a flaw that becomes apparent only when examining permutations of messages from independently deployed user agents.

The test system comprises a vulnerable main contract and multiple user agent contracts, modeling real-world TON applications where users interact via their own contract instances, potentially deployed across different shards.

The core of the system is a main contract that manages a deposit pool using two primary state variables: a 64-bit integer `balance` and a slice \texttt{owner\_address}. Upon deployment, the balance is zero and the owner address is empty. The contract processes two operations: \texttt{OP\_ENLIST} (value 1) for deposits and \texttt{OP\_CLAIM} (value 2) for withdrawals. Each user (e.g., Alice and Bob) operates through a \texttt{UserContract} agent, which stores the main contract's address and forwards deposit requests.

The vulnerability is in the ownership assignment logic within the main contract's \texttt{OP\_ENLIST} handler. Ownership is assigned to the sender of the first deposit message. This pattern creates a race condition when multiple user agents concurrently send their first deposit messages, as their processing order is non-deterministic in TON's sharded architecture.

The following listings present the vulnerable contracts.

\textbf{Main Contract (\texttt{MainContract.fc}):}
\begin{lstlisting}[language=C++,caption={Vulnerable Main Contract with Inter-Contract Race Condition},xleftmargin=2.0ex]
#include "imports/stdlib.fc";

const OP_ENLIST = 1;
const OP_CLAIM = 2;

(int, slice) _load_state() {
  slice ds = get_data().begin_parse();
  int balance = 0;
  slice owner_address = "";
  // Load shared state: balance and owner
  if (slice_bits(ds) >= 64 + 267) {
    balance = ds~.load_uint(64);
    owner_address = ds~.load_msg_addr();
  }
  return (balance, owner_address);
}

() recv_internal(int msg_value, cell in_msg, slice in_msg_body) impure {
  var (balance, owner_address) = _load_state();
  slice sender_address = get_msg_sender(in_msg);
  if (slice_bits(in_msg_body) >= 32) {
    int op = in_msg_body~.load_uint(32);
    // DEPOSIT OPERATION: balance increment and owner assignment
    if (op == OP_ENLIST) {
      balance += msg_value;
      // VULNERABILITY: Check-then-act on shared state `owner_address`.
      // Two concurrent messages can both pass the empty check before either updates the state.
      if (slice_bits(owner_address) == 0) {
        _save_state(balance, sender_address);
      } else {
        _save_state(balance, owner_address);
      }
      return ();
    }
    // CLAIM OPERATION: withdrawal by current owner
    if (op == OP_CLAIM) {
      if (equal_slices(owner_address, sender_address)) {
        send_raw_message(begin_cell()
          .store_uint(0x18, 6)
          .store_slice(sender_address)
          .store_coins(balance)
          .store_uint(0, 1 + 4 + 4 + 64 + 32 + 1 + 1)
          .store_uint(OP_CLAIM, 32)
          .end_cell(),
          64
        );
        balance = 0;
        _save_state(balance, "");
      }
      return ();
    }
  }
}
\end{lstlisting}

\textbf{User Agent Contract (\texttt{UserContract.fc}):}
\begin{lstlisting}[language=C++,caption={User Agent Contract for Inter-Shard Communication},xleftmargin=2.0ex]
#include "imports/stdlib.fc";
const OP_SEND_ENLIST = 1;

() send_enlist(slice main_address, int amount) impure {
  cell msg = begin_cell()
    .store_uint(0, 6)                 // internal, no bounce
    .store_slice(main_address)        // MainContract address
    .store_coins(amount)              // deposit amount
    .store_uint(0, 1 + 4 + 4 + 64 + 32 + 1 + 1)
    .store_uint(OP_ENLIST, 32)        // operation for MainContract
    .end_cell();
  send_raw_message(msg, 3);           // mode 3: carry remaining balance
}

() recv_internal(int msg_value, cell in_msg, slice in_msg_body) impure {
  if (slice_bits(in_msg_body) >= 32) {
    int op = in_msg_body~.load_uint(32);
    if (op == OP_SEND_ENLIST) {
      int amount = in_msg_body~.load_uint(128);
      slice main_address = _load_main_address();
      send_enlist(main_address, amount); // Cross-contract invocation
      return ();
    }
  }
}
\end{lstlisting}

Consider a scenario with three contract instances:
\begin{itemize}
    \item \texttt{UserContract} for Alice (address \texttt{addr\_alice})
    \item \texttt{UserContract} for Bob (address \texttt{addr\_bob})
    \item \texttt{MainContract} (address \texttt{addr\_main})
\end{itemize}

Both Alice and Bob simultaneously initiate their first deposits via their agents, generating the following messages to the main contract:
\begin{enumerate}
    \item \texttt{OP\_ENLIST} from \texttt{addr\_alice}
    \item \texttt{OP\_ENLIST} from \texttt{addr\_bob}
    \item \texttt{OP\_CLAIM} from \texttt{addr\_alice} (attempting to withdraw the pooled balance)
\end{enumerate}

Due to TON's asynchronous architecture, the processing order of messages 1 and 2 is non-deterministic, leading to divergent financial outcomes:

\textbf{Scenario A: Message 1 (Alice) processes first.}
\begin{enumerate}
    \item Alice's deposit executes. \texttt{owner\_address} is empty, so Alice becomes the owner.
    \item Bob's deposit executes. \texttt{owner\_address} is now non-empty, so Bob is not set as owner.
    \item Alice's claim executes. Ownership verification succeeds; Alice withdraws the full balance.
    \item \textbf{Final state}: Balance = 0, Owner = (empty).
\end{enumerate}

\textbf{Scenario B: Message 2 (Bob) processes first.}
\begin{enumerate}
    \item Bob's deposit executes. \texttt{owner\_address} is empty, so Bob becomes the owner.
    \item Alice's deposit executes. \texttt{owner\_address} is now non-empty; ownership remains with Bob.
    \item Alice's claim executes. Ownership verification fails; the withdrawal is rejected.
    \item \textbf{Final state}: Balance = (total deposits), Owner = \texttt{addr\_bob}. Alice loses her funds.
\end{enumerate}

This is a classic check-then-act race condition. The check \texttt{slice\_bits(owner\_address) == 0} and the subsequent state update are not atomic. Concurrent messages from different senders can interleave between these steps, causing the contract to assign ownership based on a transient, outdated state view.

The vulnerability is subtle because:
\begin{itemize}
    \item It passes linear, single-user functional testing.
    \item It manifests only under specific message orderings.
    \item It results in a complete, silent loss of funds for one party.
    \item It is a direct consequence of TON's non-deterministic inter-shard message ordering.
\end{itemize}

BugMagnifier systematically uncovers this flaw through permutation testing. For the three-message sequence, it explores all $3! = 6$ possible orderings. The \texttt{diff} analysis reveals two distinct final state classes corresponding to the two scenarios above, clearly identifying the non-deterministic outcome and the critical divergence in the \texttt{owner\_address} field. This case study exemplifies how asynchronous execution transforms simple state transitions into critical vulnerabilities, necessitating tools like BugMagnifier for detection.

\subsection{Real-World Vulnerability Case Studies}
\label{subsec:real-world-cases}

To validate the practical applicability of BugMagnifier beyond synthetic benchmarks, we reproduced five critical vulnerability patterns identified in recent TON smart contract security audits. These cases span diverse vulnerability classes including race conditions, improper bounce handling, state consistency errors, and authorization flaws. Each case was implemented in FunC and tested using the BugMagnifier framework to confirm detectability through message permutation and state differential analysis.

Table~\ref{tab:case-studies} summarizes the reproduced vulnerabilities. The MeMeStore case demonstrates a race condition in which concurrent transactions skip a trade step threshold, leading to fund loss. The Aqua Protocol case highlights improper state cleanup during bounce processing, causing operational deadlock. ThunderFinance illustrates a state consistency issue where balance updates precede external calls without bounce handlers, resulting in permanent fund loss. The MAI1-1 case reveals an authorization flaw in which sender validation occurs after state-changing logic, allowing unauthorized refunds in paused states. Finally, the EVAA re-activation race condition~\cite{Quantstamp2024EVAA} exhibits unfair competition between borrowers and liquidators upon system resumption; this pattern was identified in audit reports but excluded from empirical evaluation due to contract unavailability.

\begin{table*}[t]
\centering
\caption{Real-World Vulnerability Case Studies Reproduced and Validated}
\label{tab:case-studies}
\begin{tabularx}{\textwidth}{@{}>{\raggedright\arraybackslash}p{2.5cm} >{\raggedright\arraybackslash}p{3cm} >{\raggedright\arraybackslash}p{12cm}@{}}
\toprule
\textbf{Project} & \textbf{Vulnerability Class} & \textbf{Description and BugMagnifier Detection} \\
\midrule
\textbf{MeMeStore}~\cite{TonBit2024MeMe} & Race Condition & Concurrent transactions skip trade step threshold. BugMagnifier detects divergent final balances when \texttt{ENLIST} messages interleave before state update. \\
\addlinespace
\textbf{Aqua Protocol}~\cite{Beosin2024Aqua} & Bounce Handling & Multi-step redeem data sticks on bounce. Tool identifies state divergence when \texttt{recv\_bounced} fails to clear context for all operation types. \\
\addlinespace
\textbf{ThunderFinance}~\cite{TonBit2024Thunder} & State Consistency & State updates before external call without bounce handler. Detection via balance mismatch after failed external message returns. \\
\addlinespace
\textbf{MAI1-1 (Torch)}~\cite{TonBit2024Torch} & Authorization Order & Delayed sender check allows refund in paused state. Tool flags unauthorized state transition when message order bypasses validation logic. \\
\bottomrule
\end{tabularx}
\end{table*}

\begin{figure*}[t]
\centering
\includegraphics[width=0.75\linewidth]{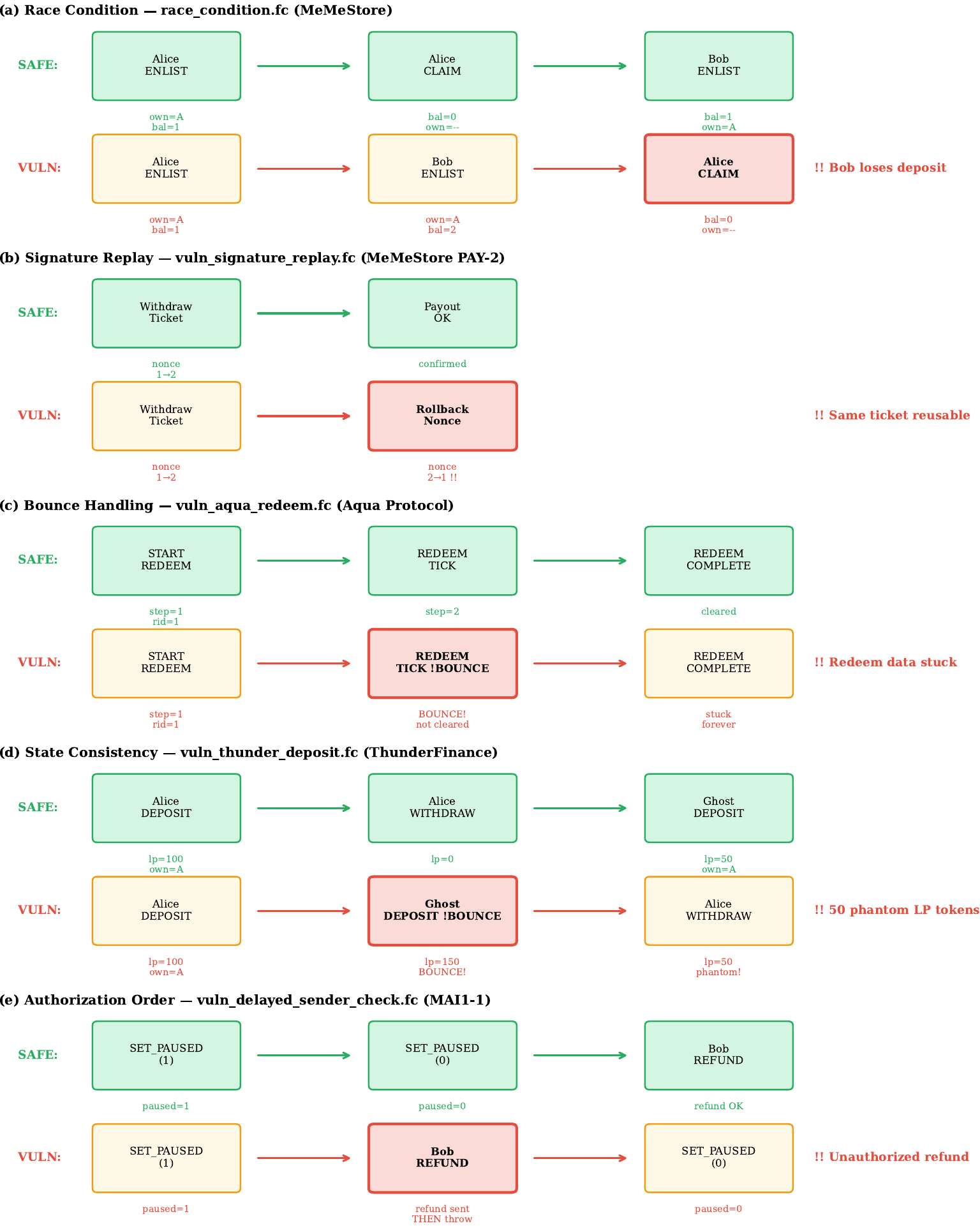}
\caption{Message queue ordering impact on five vulnerability classes. Each panel contrasts safe execution (left) with vulnerable interleaving (right). (a) Race condition: concurrent \texttt{ENLIST} messages observe empty \texttt{owner\_address}. (b) Signature replay: nonce rollback enables ticket reuse. (c) Bounce handling: failed redeem step leaves state uncleared. (d) State consistency: balance update precedes external call without bounce handler. (e) Authorization order: sender validation occurs after refund logic.}
\label{fig:vulnerability-flows}
\end{figure*}

Figure~\ref{fig:vulnerability-flows} illustrates the message ordering patterns that trigger each vulnerability class. For each case, the safe execution path is contrasted with the vulnerable interleaving that leads to state divergence. The diagrams show how specific message sequences produce divergent contract states, confirming that BugMagnifier's permutation testing captures the temporal dependencies underlying these real-world flaws.

For each case, BugMagnifier successfully identified the vulnerability by exploring message orderings that static analyzers typically miss. For instance, in the ThunderFinance scenario, the tool detected that when a deposit message triggers an external call that fails due to insufficient gas, the lack of a bounce handler leaves the contract state inconsistent (increased supply without received funds). By simulating the bounce message explicitly, the framework confirmed the state divergence. Similarly, for the MAI1-1 authorization flaw, randomizing the message queue revealed execution paths where the paused state check was bypassed by interleaving administrative and user messages. These results confirm that the temporal vulnerabilities modeled in our synthetic experiments (Section~\ref{sec:NumericalExperiments}) accurately reflect challenges present in production TON smart contracts.

\section{Numerical Experiments}
\label{sec:NumericalExperiments}

To evaluate BugMagnifier's effectiveness, we conducted two complementary sets of experiments. First, we used the synthetic race condition contract from Section~\ref{sec:Example} to establish a quantitative model of detection complexity under controlled message ratios. Second, we validated this model by applying BugMagnifier to the five real‑world vulnerability cases reproduced from recent security audits. This dual approach demonstrates both the statistical predictability of race condition manifestation and the practical applicability of our framework to production‑grade flaws.

All experiments were conducted using BugMagnifier~\cite{BugMagnifier} on a local TON Sandbox instance. Message queues were generated from audit report specifications and production traffic estimates. Each configuration was executed 100 times with randomized message ordering; state divergence was detected via cell-level comparison of contract storage and get-method outputs. Source code and test vectors are available in the public repository~\cite{BugMagnifierInAction}.

\subsection{Synthetic Race Condition Experiments}
\label{subsec:synthetic-experiments}

The experimental framework generated message queues containing $n_1$ messages from Alice and $n_2$ messages from Bob, structured to model the race condition patterns observed in the real-world case studies. While the real-world contracts involve complex logic, the core temporal vulnerability mechanism reduces to competing message sequences affecting shared state. Each configuration was executed using BugMagnifier with the command line interface described in Section~\ref{sec:ProposedSolution}. The experimental protocol executed 1,000 iterations per configuration. In each iteration, the message queue was randomly shuffled and processed sequentially. After the first iteration, the initial state was saved, and subsequent iterations began with a reset environment. The state divergence between the current execution and baseline, detected through internal state comparison and get-method results, terminated the iteration sequence.

The methodology evaluated race condition detection complexity across message ratios with fixed Bob messages ($n_2 = 32$) and geometrically increasing Alice messages ($n_1 = 1, 2, 4, \dots, 512$). For each configuration, 100 independent trials recorded the iteration count where state divergence first occurred, providing a quantitative measure of a contract's susceptibility to race conditions.

\begin{figure}[h]
	\centering
	\includegraphics[width=0.95\columnwidth]{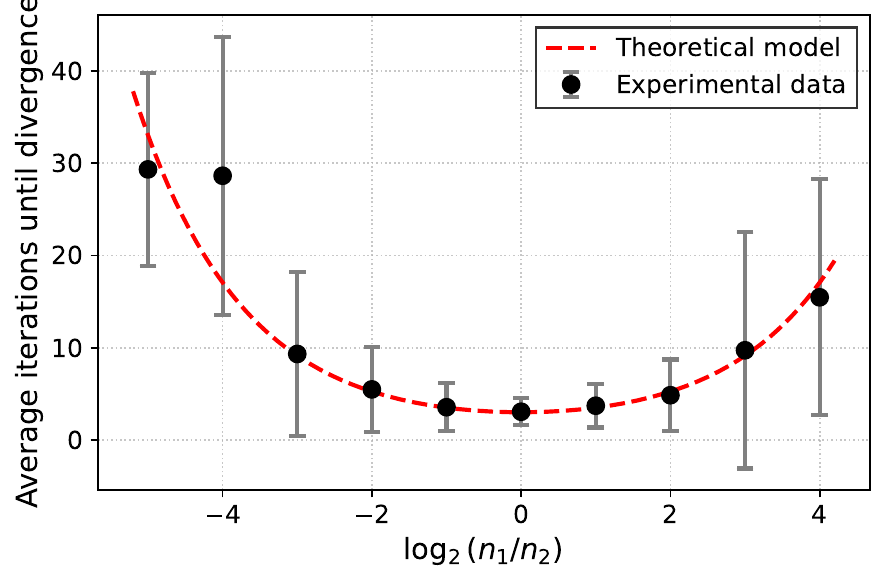}
	\caption{Analysis of race condition in a smart contract: average number of iterations until first state divergence versus message ratio $\log_2(n_1/n_2)$, with $n_2 = 32$ fixed and $n_1$ varied geometrically. Results are based on 100 independent launches per configuration. Error bars denote $\pm1$ standard deviation.}
	\label{fig:graphic}
\end{figure}

Figure~\ref{fig:graphic} plots the results against $\log_2(n_1/n_2)$, with error bars representing $\pm1$ standard deviation. The data reveal a clear U‑shaped relationship between the message ratio and detection complexity. Minimum detection iterations (3.06) occur at balanced message counts ($n_1 = n_2 = 32$), while extreme imbalances require significantly more iterations (29.32 for $n_1 = 1$, 15.46 for $n_1 = 512$). Higher standard deviations at imbalanced ratios indicate greater variance in detection timing. The theoretical curve, derived from the expectation formula where $p$ and $q$ are the relative message probabilities, aligns precisely with the experimental points, confirming the probabilistic model of race condition manifestation.

\subsection{Empirical Validation on Real-World Cases}
\label{subsec:real-world-validation}

To assess the generality of our quantitative model, we applied BugMagnifier to the real-world vulnerability patterns described in Section~\ref{subsec:real-world-cases}. For each case with an available contract implementation, we constructed message queues reflecting the operational sequences specified in the corresponding audit reports. Table~\ref{tab:real-world-metrics} summarizes the detection metrics for four implemented cases.

The MeMeStore race condition was detected after an average of 1.86 iterations (std.\ dev.\ 1.32) under a 2:1 message ratio (two \texttt{ENLIST}, one \texttt{CLAIM}). For the minimal 3-message scenario, exhaustive enumeration shows 2 of 6 permutations produce vulnerable final states ($f = 1/3$), yielding a theoretical detection expectation of $\mathbb{E}[X] = 3.25$ iterations under random baseline sampling. The observed lower average (1.86) suggests that, in practice, the effective vulnerable fraction increases due to additional interleaving patterns present in the full contract logic and production-like message ratios. The Aqua Protocol bounce handling flaw required 2.13 iterations on average (std.\ dev.\ 1.51) under a 1:2 ratio (START, COMPLETE, TICK), reflecting the need to interleave a bounce message with the redeem sequence. ThunderFinance exhibited a detection complexity of 3.04 iterations (std.\ dev.\ 2.16) under a 2:1 deposit/withdrawal ratio, corresponding to the state update preceding the external call. The MAI1-1 authorization flaw in Torch required the most iterations among implemented cases (6.14, std.\ dev.\ 5.03) under a 2:1 administrative/user message ratio, as the vulnerability manifests only when the paused-state check is bypassed by specific interleavings.

One additional vulnerability pattern from our case studies warrants discussion despite its exclusion from the quantitative table. The signature replay vulnerability (\texttt{vuln\_signature\_replay}) was detected deterministically in a single iteration, as the nonce rollback mechanism creates an immediate state divergence upon processing the \texttt{RollbackNonce} message; this trivial detection complexity reflects the vulnerability's structural simplicity rather than a limitation of our methodology, and was therefore omitted from Table~\ref{tab:real-world-metrics} to avoid skewing the statistical summary.

These results confirm that temporal vulnerabilities in production TON smart contracts can be detected with low iteration counts when message queues reflect realistic operational ratios. The observed detection complexities align with the theoretical model established in Section~\ref{subsec:synthetic-experiments}: vulnerabilities with smaller message spaces and more constrained interleaving requirements are detected more rapidly. This finding supports targeted testing strategies informed by expected traffic patterns, rather than exhaustive permutation exploration. By quantifying detection effort for real-world audit scenarios, BugMagnifier provides a predictive model for security testing resource allocation.

\begin{table*}[h]
\centering
\footnotesize
\setlength{\tabcolsep}{4pt}
\caption{Detection Metrics for Real-World Vulnerability Cases}
\label{tab:real-world-metrics}
\begin{tabular}{|l|r|r|l|l|}
\hline
\textbf{Project} & \textbf{Avg. Iter.} & \textbf{Std. Dev.} & \textbf{Msg Ratio} & \textbf{Vulnerability Class} \\
\hline
MeMeStore~\cite{TonBit2024MeMe} & 1.86 & 1.32 & 2:1 (2 ENLIST, 1 CLAIM) & Race Condition \\
\hline
Aqua Protocol~\cite{Beosin2024Aqua} & 2.13 & 1.51 & 1:2 (START, COMPLETE, TICK) & Bounce Handling \\
\hline
ThunderFinance~\cite{TonBit2024Thunder} & 3.04 & 2.16 & 2:1 (2 DEPOSIT, 1 WITHDRAW) & State Consistency \\
\hline
MAI1-1 (Torch)~\cite{TonBit2024Torch} & 6.14 & 5.03 & 2:1 (PAUSE, REFUND, UNPAUSE) & Authorization Order \\
\hline
\end{tabular}
\end{table*}

\subsection{Threats to Validity}
\label{subsec:threats-to-validity}

Whilst BugMagnifier provides systematic detection of temporal vulnerabilities, several limitations warrant consideration. First, the fidelity of TON Sandbox emulation, though high, may not capture all mainnet conditions, particularly network-level latency variations and validator-specific gas scheduling that could affect message interleaving probabilities. Second, our message queue modeling assumes independent, uniformly random ordering; real-world traffic patterns may exhibit temporal correlations or adversarial scheduling that alter vulnerability manifestation rates. Third, the permutation testing approach, while effective for small message sets, faces combinatorial explosion for large queues, necessitating heuristic sampling that may miss rare but critical orderings. Finally, the synthetic contracts used for quantitative modeling abstract away complex business logic present in production systems, potentially underestimating detection complexity for contracts with intricate state dependencies. These limitations suggest directions for future work, including integration with mainnet traffic traces, adaptive sampling strategies, and hybrid analysis combining permutation testing with symbolic execution to scale to larger message spaces.

\section{Conclusion}  
\label{sec:Conclusion}  

Our research establishes BugMagnifier as a foundational advancement in addressing the unique security challenges posed by TON's asynchronous execution model. Unlike conventional blockchain security tools designed for atomic transaction processing, BugMagnifier specifically targets the temporal vulnerabilities inherent to TON's message-driven architecture, particularly race conditions that emerge from non-deterministic message interleaving across the five-phase transaction lifecycle. The framework's capability to systematically explore message ordering permutations provides the first practical methodology for detecting vulnerabilities that conventional static analysis and symbolic execution approaches cannot reliably identify.

The core innovation lies in BugMagnifier's message granularity control, which enables precise manipulation of execution flows while maintaining complete visibility into TVM state transitions across all transaction phases. This capability addresses the fundamental limitation of existing tools that typically operate at the transaction level and remain blind to intermediate state changes between message processing stages. The successful reproduction of five distinct vulnerability patterns from recent security audits confirms the framework's practical utility. By transforming the detection of temporal vulnerabilities from an ad-hoc, expert-dependent process into a systematic, evidence-based methodology, BugMagnifier establishes a new paradigm for security analysis in asynchronous blockchain environments.

The implications of this work extend beyond immediate vulnerability detection. BugMagnifier's approach establishes a framework for developing security-aware design patterns specifically tailored to TON's architectural constraints, potentially influencing future smart contract development practices within the ecosystem. The methodology also provides a template for addressing similar challenges in other asynchronous blockchain platforms, suggesting a broader applicability of the core concepts.

Looking forward, this research opens several promising directions. The integration of BugMagnifier's permutation testing methodology with formal verification techniques could yield hybrid analysis approaches with stronger guarantees. Additionally, the message scheduling capabilities of the framework could inform the development of defensive programming patterns that inherently resist temporal vulnerabilities. As TON continues to mature, tools like BugMagnifier will play a critical role in establishing security standards that match the ecosystem's technical sophistication, ultimately enabling more robust and reliable decentralized applications on the platform.


\bibliographystyle{elsarticle-num}
\bibliography{refs2025}

\end{document}